\documentclass[5p,twocolumn,times,number]{elsarticle}

\usepackage{graphicx}
\usepackage{amsmath}   
\def\pmbanner{{\hrule height 1 pt}\vskip35pt{NIMA POST-PROCESS BANNER TO BE REMOVED AFTER FINAL ACCEPTANCE}\vskip35pt{\hrule height 4pt}\vskip20pt}

\begin{document}

\begin{frontmatter}

\title{\pmbanner Upgrade plans and ageing studies for the CMS muon system \\ in preparation of HL-LHC}

\author{Jian Wang \\ On behalf of the CMS Collaboration}


\address{University of Florida, USA}

\begin{abstract}
The CMS muon system operates gas-based detectors. Upgrades of the detectors and trigger components are needed to cope with increasingly challenging conditions of the HL-LHC. New irradiation tests are performed to ensure that the muon detectors will survive the harsher conditions and operate reliably. The new CERN GIF++ (Gamma Irradiation Facility) allowed to perform aging tests. We present results in terms of system performance under large backgrounds and after accelerated aging tests. New detectors will be added to improve the performance in the critical forward region: gas electron multiplier (GEM) detectors will be installed in LS2 in the region 1.6 $<$ $|\eta|$ $<$ 2.4, aiming at suppressing background triggers while maintaining high trigger efficiency. Further enhancements are foreseen with a second GEM station and with two stations of new generation RPCs, having low resistivity electrodes. These detectors will combine tracking and triggering capabilities and can stand particle rates up to few kHz/cm$^2$. In addition, to take advantage of the pixel tracking coverage extension, a new GEM detector covering up to $|\eta|$ = 2.8. is foreseen behind the new forward calorimeter.

\end{abstract}


\end{frontmatter}

\section{Introduction}

CMS is a general purpose detector at CERN LHC \cite{Chatrchyan:2008aa}. As we know, hadrons are enormously produced at proton collisions due to QCD processes, but almost all hadrons, as well as electrons and photons, are absorbed in the calorimeters. Therefore the trigger, identification and measurement of muons is of great importance at a hadron collider experiment. The CMS muon system has been working stably and with high efficiency since 2010 \cite{Chatrchyan:2013sba,Sirunyan:2018fpa}, thus contributing crucially to the Higgs boson discovery \cite{Chatrchyan:2012xdj} as well as searches and measurements for many other physics signatures of muon final states.

\begin{figure}
\centering
\includegraphics[width=0.99\linewidth]{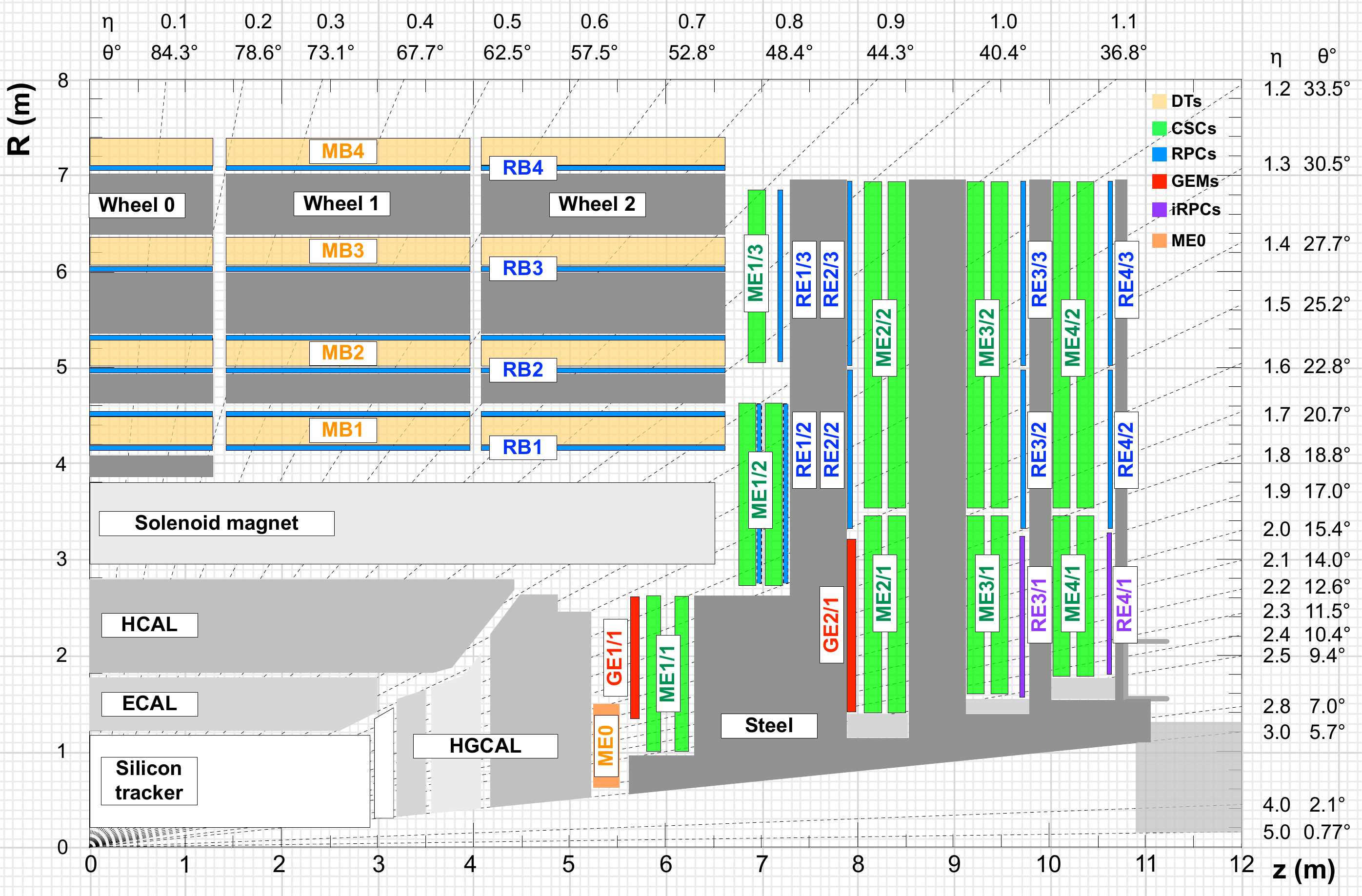}
\caption{
An $R$-$z$ cross section of a quadrant of the CMS detector, including the Phase-2
upgrades (RE3/1, RE4/1, GE1/1, GE2/1, ME0). The acronym iRPCs in the legend refers to the new
improved RPC chambers RE3/1 and RE4/1.
The interaction point is at the lower left corner. The locations of
the various muon stations are shown in color (MB = DT = Drift Tubes,
ME = CSC = Cathode Strip Chambers,
RB and RE = RPC = Resistive Plate Chambers, GE and ME0 = GEM = Gas Electron Multiplier).
M denotes Muon, B stands for Barrel and E for Endcap.
The magnet yoke is represented by the dark gray areas.
}
\label{fig:muon_detector}
\end{figure}

The muon system is the outmost part of the CMS detector (Figure \ref{fig:muon_detector}). It sits outside the 3.8~T superconducting solenoid magnet, and sandwiched between steel plates of return yoke. The present  muon system consists of the barrel part and the endcap part, and both parts have 4 "stations". Three types of gas detectors are used. Drift Tubes (DT) are chosen in the barrel ($|\eta|$ $<$ 1.2, where $\eta$ is pseudorapidity) and Cathode Strip Chambers (CSC) cover the endcap (0.9 $<$ $|\eta|$ $<$ 2.4). Different detector technologies are chosen for the reason that the particle rates in different $\eta$ regions could be different by more than a factor of 100, with the higher $\eta$ region larger. (Another factor driving detector technology choice is the magnet field). Both DT and CSC have very good spatial resolution (100~$\mu$m or better), but DT does not operate well under high particle rate and high magnetic filed, due to larger drift path. In both barrel and endcap, Resistive Plate Chambers (RPC) supplement DT or CSC and work as trigger chambers, taking advantage of its good timing resolution. CSC chambers have 6 layers (6 hit measurements per chamber on the path of a muon). DT chambers have 8 layers to measure the coordinate where muons bend in the magnetic filed, and another 4 layers the orthogonal coordinate. RPC chambers have 2 layers in the first 2 barrel stations, and 1 layer elsewhere. If a muon comes out of the collision point, given sufficient momentum, it passes 4 stations and 2 types of muon detectors, which guarantees robust trigger and efficient reconstruction.

The present muon system was designed to operate for the initial LHC specifications with the instantaneous luminosity up to 10$^{34}$cm$^{-2}$s$^{-1}$. A major upgrade of the LHC is being prepared, referred to as the High Luminosity LHC (HL-LHC) \cite{Apollinari:2015bam,ApollinariG.:2017ojx}. The high luminosity data taking period will follow a shutdown for the machine upgrade finishing in 2026, and last for about 10 years. The integrated luminosity is expected to reach 3000 fb$^{-1}$, one order of magnitude higher with respect to the original LHC design value. It immediately raises the question if the present muon detectors are sufficiently radiation hard. 
On the other hand, some muon system electronics replacement is inevitable, required by higher latency and bandwidth as the HL-LHC provides instantaneously luminosity of 5-7 $\times$ 10$^{34}$ cm$^{-2}$s$^{-1}$, much higher than that of today. (Electronics replacement will not be detailed in this report due to limited space.)

\section{Muon detector ageing studies}

As the HL-LHC provides radiation levels well beyond the LHC design specification, new radiation tests must be performed to confirm that existing muon detectors will survive. Exposure to high radiation could potentially cause detector deterioration and permanent failure. The symptoms could be gas gain decrease, spurious hits, self-sustained discharges, HV breakdown and so on. 

To certify muon detector longevity at the HL-LHC level, a new Gamma Irradiation Facility (GIF++) \cite{Pfeiffer:2016hnl} has been built at CERN. At GIF++, detectors and material samples can be irradiated by 662~keV photons emitted by an intense 13.5~TBq Cs-137 source. Full-size DT, CSC and RPC chambers are exposed to high rates here at GIF++. In such accelerated irradiation,  the accumulated charge per cm of wire or cm$^2$ area is the proxy of "radiation exposure". The HV currents observed in present CMS chambers are used to extrapolate to the required total accumulated charge corresponding to HL-LHC. A safety factor of 3 is required in addition.

\begin{figure}
\centering
\includegraphics[width=0.99\linewidth]{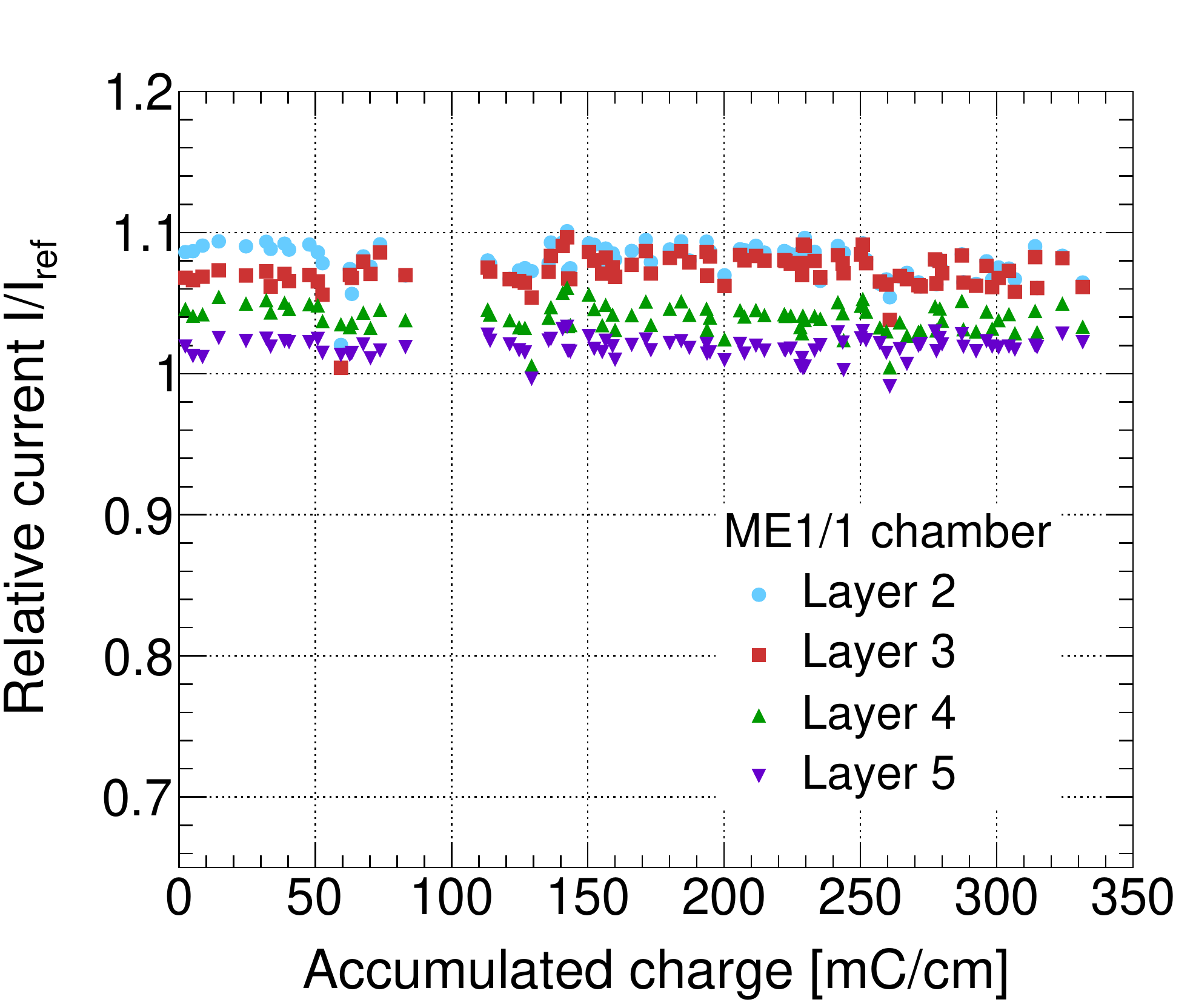}
\caption{
Relative currents normalized to reference layers that are turned off during irradiation test, as a function of the total accumulated charge for a CSC ME1/1 chamber.
}
\label{fig:CSCgain}
\end{figure}

Chamber performance is monitored as a function of the accumulated charge. CSC and RPC have finished their accelerated irradiation test, and no noticeable performance degradation has been seen up to 3 times HL-LHC equivalent accumulated charge. Figure \ref{fig:CSCgain} shows the  HV current of the irradiated CSC chambers, relative to some layers staying off as reference channels. No decease of the relative current is seen up to 330~mC/cm wire accumulated charge. (The HL-LHC equivalent accumulated charge is 110~mC/cm)

As for DT, about 15\% of chambers are predicted to see noticeable gas gain decrease in HL-LHC lifetime, according to the ageing test. But the muon reconstruction efficiency in the barrel is not expected to degrade significantly, thanks to multiple layers of DT on the path of a muon. Meanwhile, mitigation measures are being implemented, including reducing gas recirculation, HV adjustment, adding shielding for chambers, etc.

In summary, the recent ageing studies conclude that all existing muon chambers will not be replaced, and are expected to operate to the end of HL-LHC lifetime.

\section{New detectors in the high $\eta$ region}

The high $\eta$ region is very challenging for muon trigger and reconstruction. The background rates are the highest due to random hits, hadron punch-though as well as muons from the collision point. On the other hand, the bending of muon trajectory is small, making the $p_T$ measurement less easy. Despite these difficulties, the high $\eta$ region has fewer number of hit measurements as of today. As mentioned, 1.8 $<$ $|\eta|$ $<$ 2.4 is only equiped by CSC.

As part of the CMS muon system upgrade project, the number of hit measurements on the muon path in the high $\eta$ region is going to be increased by adding new muon detectors. The locations of the additional detectors, specifically ME0, GE1/1, GE2/1, RE3/1, RE4/1 are indicated in Figure \ref{fig:muon_detector}.

At endcap muon stations 3 and 4, new RPC chambers with improved performance, iRPC for short, are going to extend the RPC coverage from $|\eta|$ = 1.9 to 2.4, next to the existing ME3/1 and ME4/1 CSC chambers. The iRPC consists of double-gap RPC units, same as the present CMS RPC. But the gas gap is 1.4~mm instead of 2~mm. The operational HV is lowered. The loss in gas gain will be compensated by higher signal amplification of front-end electronics. It has been extensively tested at GIF++ and found to handle well particle rate of 2~kHz/cm$^2$, 3 times of the expected HL-LHC rate at the region where iRPC is going to be installed.

\begin{figure}
\centering
\includegraphics[width=0.44\linewidth]{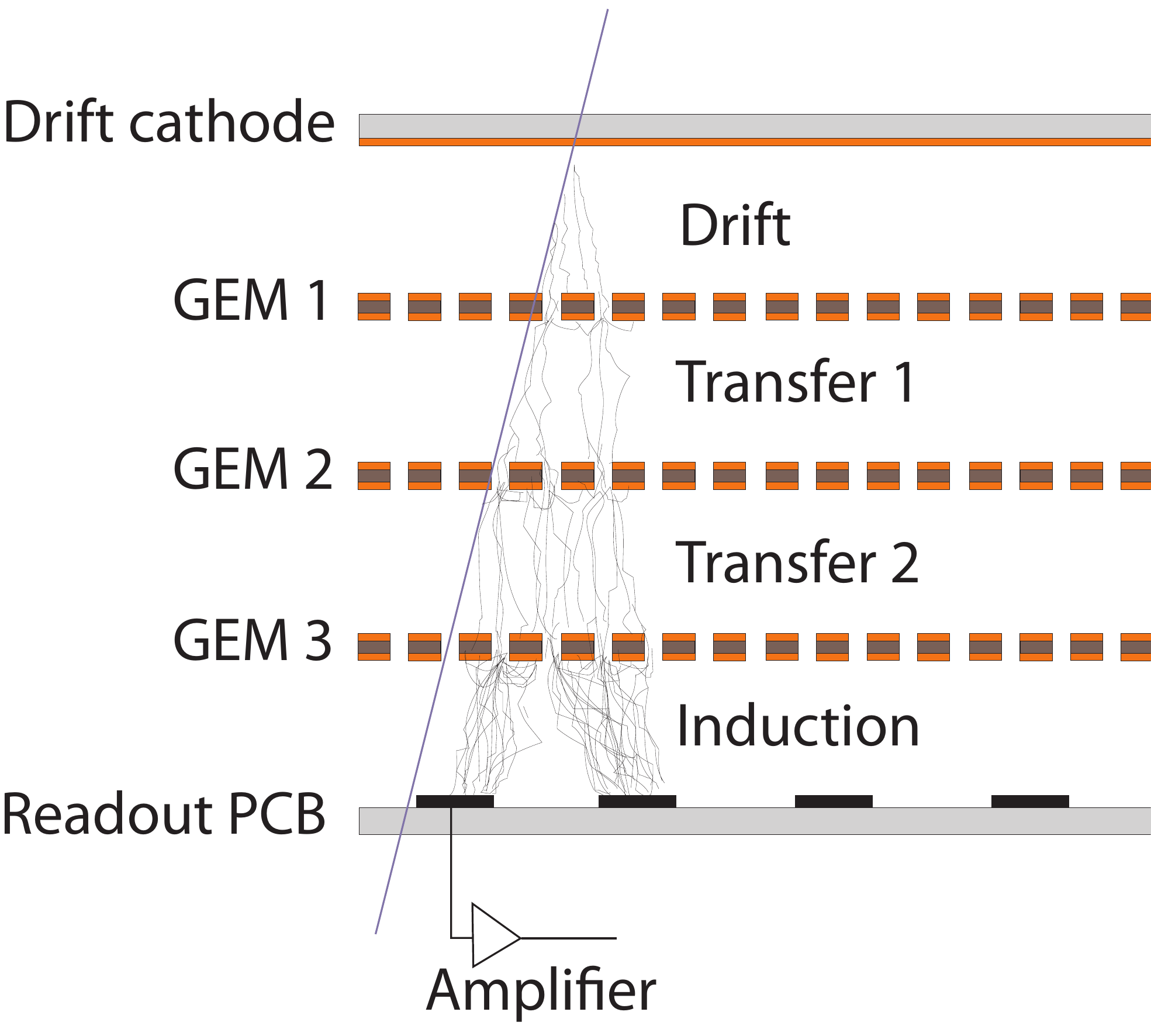}
\includegraphics[width=0.54\linewidth]{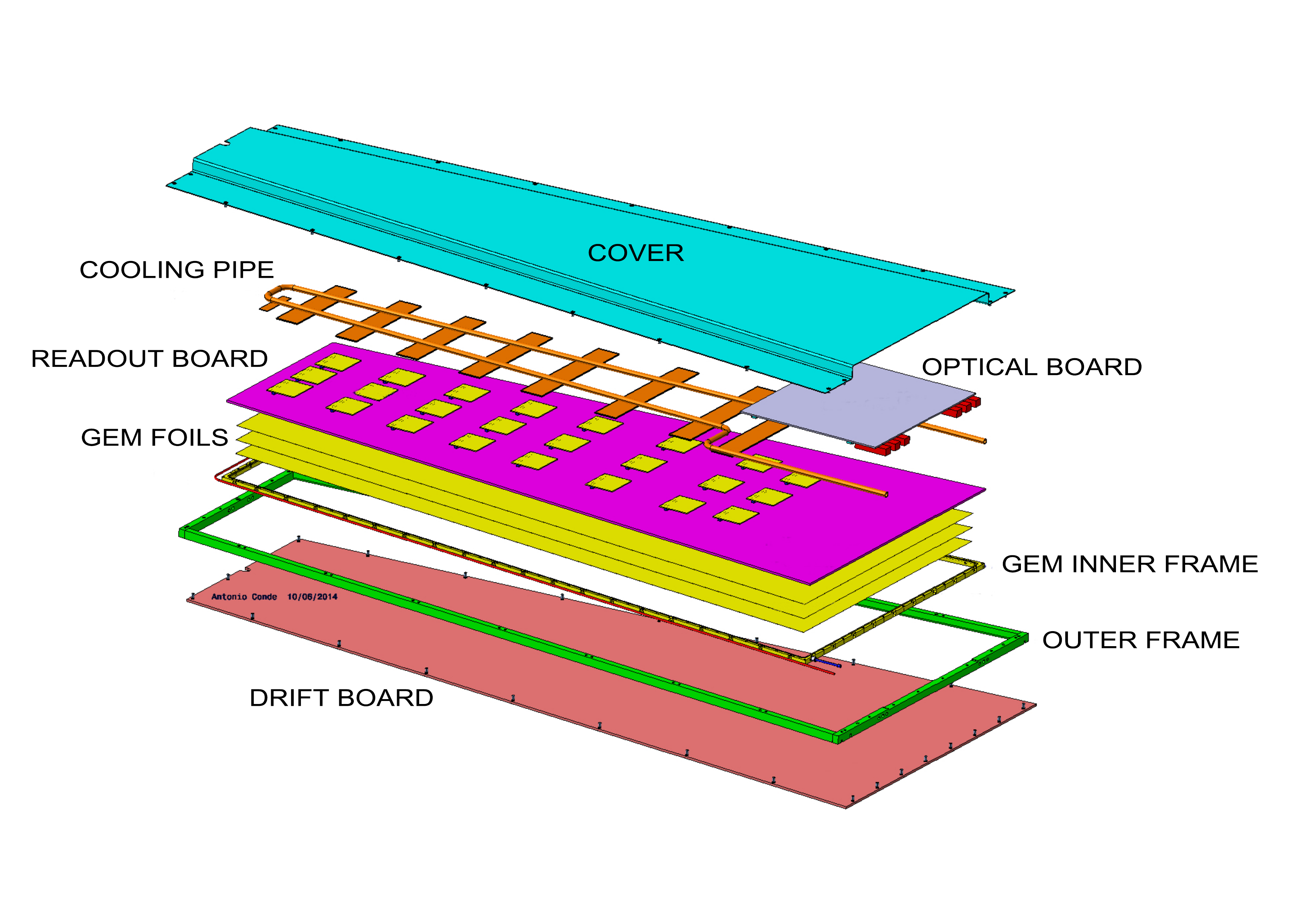}
\caption{
Left: arrangement of the triple-GEM chamber with three foils; the drift electrode on the top and the readout electrode at the bottom define drift and induction fields. Combing three stages with a gas gain 20 each results in a total amplification factor of 8000. Right: exploded view of the mechanical design of a triple-GEM chamber.
}
\label{fig:GEMlayout}
\end{figure}

At the first two endcap muon stations, Gas Electron Multiplier (GEM) detectors GE1/1 and GE2/1 are proposed to complement the existing ME1/1 and ME2/1 CSC chambers. The GEM detector makes use of avalanches in strong electric field concentrated in pin holes. Triple-GEM design is chosen such that gas gain of $10^4$ can be achieved if each GEM foil has a gain of 20-25. Layout of the triple-GEM design is shown in Figure \ref{fig:GEMlayout}. GEM has proven to operate reliably at particle rates of the order of a few MHz/cm$^2$, and is known to have excellent longevity. The spatial resolution is about 100~$\mu$m, defined by the pitch between holes.
GE1/1 covers 1.6 $<$ $|\eta|$ $<$ 2.2, and GE2/1 covers 1.6 $<$ $|\eta|$ $<$ 2.4, both being two layers triple-GEM. A pilot system of 5 pairs GE1/1 chambers were inserted in CMS at the beginning of 2017 \cite{Colaleo:2015vsq}. Invaluable experience of installation and operation has been gathered. 

A high $\eta$ muon tagger, referred to as ME0, is set to be installed behind the future endcap calorimeter. The same triple-GEM technology will be used for ME0. The coverage extends from $|\eta|$ = 2.0 to 2.8, taking advantage of the extended acceptance of the upgraded CMS pixel detector\cite{Klein:2017nke}. Each ME0 detector consists of 6 layers, providing up to 6 hit measurements per track, and therefore capability to reject neutron-induced background, and to build muon segments for trigger as well as offline reconstruction.

\section{Physics performance}

For a muon in the high $\eta$ region, the new GE1/1 and GE2/1 detectors each provide 2 additional hit measurements, and the 2 iRPC detectors add 1 hit each, while the ME0 detectors could provide 6 hits, depending on the $\eta$. As a result, the trigger and reconstuction efficiency and robustness will be improved, despite of the harsh HL-LHC environment.

\begin{figure}
\centering
\includegraphics[width=0.99\linewidth]{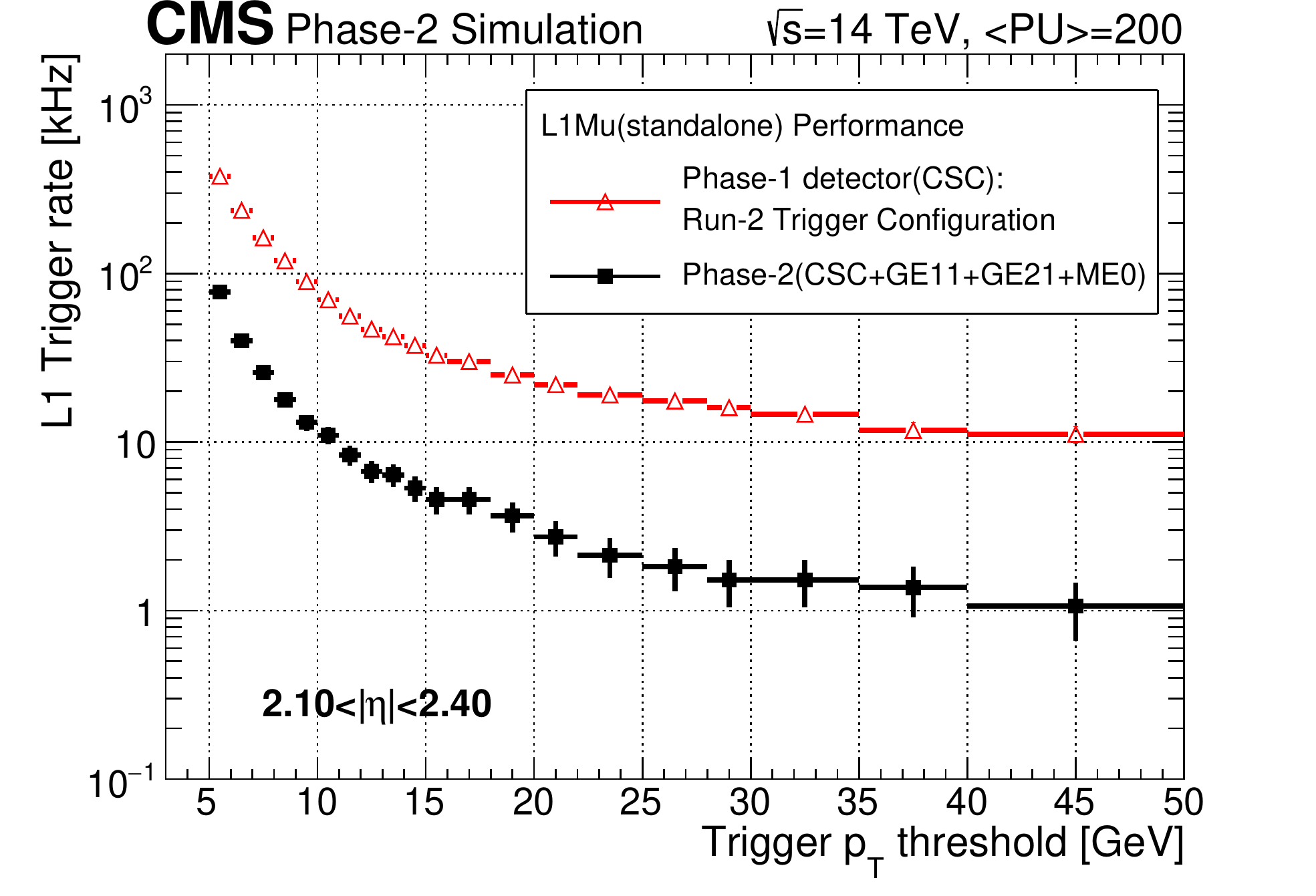}
\caption{
L1 muon trigger rates, with and without GEM chambers, as a fuction of muon trigger $p_T$ threshold in the region 2.1 $<$ $|\eta|$ $<$ 2.4. The trigger rate is expected to be reduced by a factor of 5-10 by adding GEM chambers.}
\label{fig:CSCGEM}
\end{figure}

On the other hand it is important to keep the trigger rate under control without raising trigger momentum thresholds. By having CSC-GEM tandem, the path length within a muon station is largely increased, allowing much better muon local direction measurements. This consequently improves trigger level muon momentum determination. As the background has a steeply falling momentum spectrum, better momentum resolution translates into trigger rate reduction. In fact, simulation studies show that the trigger rate is reduced by a factor of  5-10 by adding GEM detectors, as shown in Figure \ref{fig:CSCGEM}.

Adding GEM detectors also makes it possible to build trigger level muons without assuming them coming out from the collision point, enabling triggering on highly displaced muons. Meanwhile, the upgraded RPC electronics fully exploit the RPC time resolution, which allows better suppression of out-of-time background, and also allows to identify patterns of delayed hits from one muon station to the next. The latter is important for triggering on Heavy Stable Charge Particles.

\begin{figure}
\centering
\includegraphics[width=0.99\linewidth]{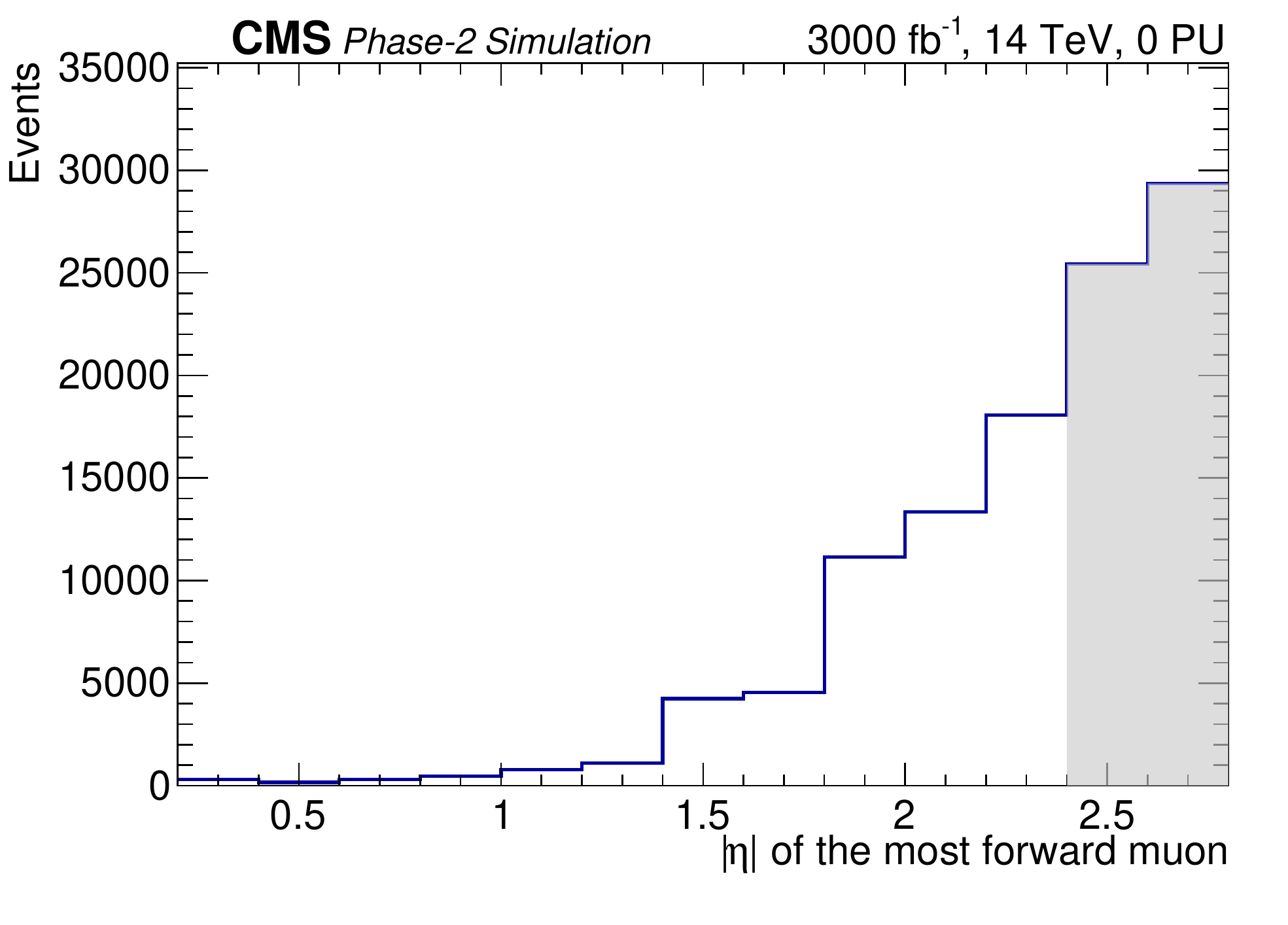}
\caption{
$|\eta|$ distribution of the most forward muon in the reconstructed $\tau$ $\rightarrow$ 3$\mu$ events. The shaded area corresponds to the $|\eta|$ range covered by ME0 chambers only.
}
\label{fig:EtaMax}
\end{figure}

A number of physics channels benefit from extended muon $\eta$ acceptance. Lepton favour violating $\tau$ $\rightarrow$ 3$\mu$ search is one of them. As $\tau$-leptons produced at LHC are boosted to the high $\eta$ region, the signal acceptance will be doubled by adding ME0 detectors (Figure \ref{fig:EtaMax}). A full simulation study taking into account the muon momentum resolution and background at 200 pile-up collision condition shows that the searching sensitivity increases 17\% compared to without ME0 detectors. Other physics analyses gaining from muon $\eta$ acceptance extension include double-parton scattering study in $pp$ $\rightarrow$ $W^+W^+$ process, electroweak mixing angle measurement with DY $\mu^+\mu^-$ events, etc.

\section{Summary}

The CMS muon system upgrade plan consists of several aspects: the longevity of existing DT, CSC, RPC chambers has been studied, and all chambers are going to stay for the lifetime of HL-LHC; various electronics components will be replaced to cope with much increased trigger and readout rate; the challenging high $\eta$ region is set to be enhanced with additional iRPC, GEM and ME0 detectors. At the same time, the upgraded detector capabilities open windows for new physics opportunities. CMS has gone though an extensive R\&D programme to optimize the upgraded detector performance while keeping the cost the lowest possible. The CMS muon detector upgrade Technical Design Reported, specifying technical details, is published\cite{CMS:muon2018}. The CSC electronics replacement and GE1/1 chambers installation will be completed in the LHC Long Shutdown 2 (2019-2020). The rest will continue in Year End Technical Stops, and finally finish in the Long Shutdown 3 (2024-mid 2026).

\section*{Acknowledgments}


\end{document}